\documentclass[11pt,UKenglish]{article}
\usepackage{xfrac}
\usepackage{tikz-cd}
\usetikzlibrary{matrix}
\usepackage{setspace}
\usepackage{xspace}
\usepackage{amsmath,amssymb,enumerate,amsthm}
\usepackage{yhmath}
\usepackage{hyperref}
\usepackage[margin = 2.50cm]{geometry}

\allowdisplaybreaks

\title{Uniform Brackets, Containers, and Combinatorial\\
Macbeath Regions}

\author{
Kunal Dutta
\footnote{Department of Informatics, University of Warsaw, Poland}
\thanks{Supported by the Polish NCN SONATA Grant no. 2019/35/D/ST6/04525.}
\and 
Arijit Ghosh
\footnote{Indian Statistical Institute}
\and
Shay Moran
\footnote{Technion and Google Research}
}

\date{}

\theoremstyle{plain}
\newtheorem{theorem}{Theorem}[section]
\newtheorem{lemma}[theorem]{Lemma}
\newtheorem{claim}[theorem]{Claim}
\newtheorem{proposition}[theorem]{Proposition}
\newtheorem{corollary}[theorem]{Corollary}
\newtheorem{definition}[theorem]{Definition}

\newtheorem{remark}[theorem]{Remark}

\newcommand{\R}{{\mathbb{R}}}

\newcommand{\N}{{\mathbb{N}}}
\newcommand{\M}{{\mathbb{M}}}

\newcommand{\pth}[1]{\ensuremath{\left(#1\right)}}

\newcommand{\calM}{{\cal M}}
\newcommand{\calB}{{\cal B}}

\newcommand{\calH}{{\cal H}}
\newcommand{\calS}{{\cal S}}
\newcommand{\calC}{{\cal C}}

\newcommand{\e}{{\varepsilon}}

\newcommand{\calP}{{\mathcal{P}}}
\newcommand{\calR}{{\mathcal{R}}}
\newcommand{\calF}{{\mathcal{F}}}

\newcommand{\calX}{{\mathcal{X}}}

\newcommand{\brac}{{N_{[\;]}}}

\begin{document}
  \maketitle

\begin{abstract}
    We study the connections between three seemingly different combinatorial structures -- \emph{uniform brackets} in statistics and probability theory, \emph{containers} in online and distributed learning theory, and \emph{combinatorial Macbeath regions}, or \emph{Mnets} in discrete and computational geometry. We show that these three concepts are manifestations of a single combinatorial property that can be expressed under a unified framework along the lines of Vapnik-Chervonenkis type theory for uniform convergence. These new connections help us to bring tools from discrete and computational geometry to prove improved bounds for these objects. Our improved bounds help to get an optimal algorithm for distributed learning of halfspaces, an improved algorithm for the distributed convex set disjointness problem, and improved regret bounds for online algorithms against $\sigma$-smoothed adversary for a large class of semi-algebraic threshold functions.\\
    
\end{abstract}





\section{Introduction}
\label{sec:intro}

       A particularly pleasing situation in theoretical studies is when seemingly independent notions arising in disparate areas with different  
       applications and techniques, turn out to have a common theoretical basis. In this article, we study a combinatorial notion whose manifestations
       appear in three different areas as distinct combinatorial objects -- as \emph{uniform brackets} in statistical learning and empirical process theory,
       \emph{containers} in online and distributed learning theory, and \emph{Combinatorial Macbeath regions}, or \emph{Mnets} in discrete and computational geometry -- and show that these are 
       consequences of an underlying combinatorial property. The close connection between uniform brackets and containers has been known~\cite{BravermanKMS21,HaghtalabRS20}. 
       We connect these notions with Mnets, which are discrete analogues of a classical theorem of Macbeath in convex geometry. This allows us to import tools from 
       discrete and computational geometry to solve problems and improve bounds in each of these areas, in some cases proving optimal new bounds.

       As we aim to keep this paper accessible to readers from all three communities, we begin with 
       a brief introduction to the notions involved. Given a probability space $(\calX,\Omega,\mu)$, together with a family $\calH$ of measurable sets 
       in $\Omega$ and a parameter $\e  \in (0,1)$, an \emph{$\e$-uniform bracket}, or {\em $\e$-bracket} for short, for $\calH$ is a family $\calB$ of measurable sets such that for every 
       $H\in \calH$, there exist sets $A,\, B\in \calB$ with 
       $$
        A\subseteq H\subseteq B \;\; \mbox{and} \;\; \mu(B\setminus A) \leq \e.
       $$
       The \emph{$\e$-bracketing number} $\brac(\calH,\mu,\e)$ 
       of $\calH$ with 
       respect to the measure $\mu$, is the smallest possible size of an $\e$-bracket for $\calH$. The logarithm of $\brac(\calH,\mu,\e)$ is referred to as the 
       \emph{bracketing entropy}.

        For a set system $(X, \calF)$, where $X$ is finite and $\calF \subseteq 2^{X}$, a family of subsets $\calB$ of $X$ is an $\e$-bracket if for all $F \in \calF$ there exist $B^{+}$ and $B^{-}$ in $\calB$ such that 
        $$
            B^{-} \subseteq F \subseteq B^{+} \;\; \mbox{and} \;\; |B^{+} \setminus B^{-}| \leq \e|X|.
        $$

       The significance of the bracketing number in empirical process theory stems from the fact that bounds 
       on $\brac(\calH,\mu,\e)$ can be used to obtain simpler and more robust versions of uniform convergence and the law of large numbers for the corresponding families 
       of events. In particular, the proof of uniform convergence using $\e$-brackets follows directly from standard concentration inequalities together with a union bound, 
       and does not require the \emph{symmetrization} trick of Vapnik and Chervonenkis~\cite{VC71}. 
       Thus uniform convergence for families of bounded bracketing number, holds even when the point sample $X$ is generated using non-\emph{i.i.d.} processes.  
       Recently, $\e$-brackets were used by Haghtalab, Roughgarden and Shetty~\cite{HaghtalabRS20} for the smoothed analysis of online 
       and differentially private learning algorithms. For a more comprehensive introduction to these topics, we refer the reader to~\cite{adamsnobel2010,GlivenkoCantelliHandel13}.

       \emph{Containers} were recently introduced by Braverman, Kol, Moran and Saxena~\cite{BravermanKMS21} to study the communication
       complexity of distributed learning problems. The choice of the term containers was inspired by the related notion of \emph{containers for 
       independent sets} in hypergraphs~\cite{10.2307/44840935, SaxThom15}.
       Given a set system $(X,\calF)$ consisting of a ground set $X$ and a family of subsets 
       $\calF\subset 2^X$, together with a parameter $\e\in (0,1)$, an 
       $\e$-container $\calC$ is a collection of subsets of $X$ such that for every set $F\in \calF$, there exists a member $C\in \calC$ such that     
       $F\subset C$ and $|C\setminus F|\leq \e n$. A set system of points and halfspaces in $\R^d$ has a set $X$ of points in $\R^d$ and the collection 
       $\calF$ as all possible subsets of $X$ which can be generated via intersection with a halfspace in $\R^d$. Braverman, Kol, Moran and Saxena~\cite{BravermanKMS21} 
       proved a new dual version of the classical \emph{Carath\'eodory's theorem} for points and halfspaces in $\R^d$, and used it to show that 
       systems of points and halfspaces in $\R^d$ have $\e$-containers of size $O\pth{(d/\e)^d}$. This allowed them to design improved protocols for 
       bounding the communication complexity of learning problems such as distributed learning of halfspaces and distributed linear programming.

       A classical theorem of Macbeath~\cite{M52} in convex geometry states that for any $\e \in (0,1)$, every convex body in $\R^d$ of unit volume 
       contains a collection of subsets, each of volume $\Omega(\e)$, 
       such that any halfspace intersecting at least an $\e$-volume of the body must contain at least one of the subsets from the collection. 
       Since its introduction Macbeath regions have been an important object of study in convex geometry~\cite{Barany00macbeathsurvey,Barany08macbeathsurvey}. More recently, Macbeath regions were used for proving data structure lower bounds~\cite{BronnimannCP93,AryaMX12}, and convex body approximation problems in computational geometry~\cite{AryaFM17,AryaFMsoda17,AryaAFMsoda20}.
       \emph{Mnets} were proposed as combinatorial analogues of Macbeath's theorem by Mustafa and Ray~\cite{MR14}, who showed their existence for many geometrically 
       defined classes of set systems. Later their result was generalized to hold for \emph{semi-algebraic} set systems with bounded \emph{shallow cell complexity} by Dutta, Ghosh, 
       Jartoux and Mustafa~\cite{DuttaGJM19}. A set system $(X,\calS)$ is said to have a $\lambda$-heavy $\e$-Mnet, if there exists a collection $\calM$ of subsets of $X$ such 
       that for any set $S\in \calS$ with at least $\e |X|$ elements, there exists a member $M$ of $\calM$ which is contained in $S$, and has at least $\lambda|S|$ elements.
       Mnets can be used to prove the existence of optimal-sized $\e$-nets for almost all studied classes of geometric set systems~\cite{DuttaGJM19}.

The rest of the paper is organized as follows. In Section~\ref{sec:rel-work-our-contribution} we present our explicit results and applications, together with some previous work. Next, 
we give some preliminary background in Section~\ref{sec:background}, followed by our general results. The proofs of our results are in Section~\ref{sec:proofs}. 
        
\section{Related work and outline of our results}
\label{sec:rel-work-our-contribution}

    Our contribution may be thought of as having two components -- a conceptual component and a technical one.
    Conceptually, our main contribution is to find the connection between three combinatorial concepts  
-- $\e$-brackets, $\e$-containers, and Mnets. Roughly, we show that the existence of any 
one of these structures in a set system implies the existence of the other two in the system or in its complement.  
To quantify these connections, we introduce the notion of Property $\M$, which essentially represents the existence of Mnets of bounded size in a set system.
These results, along with the definition of Property $\M$, are presented in Section~\ref{sec:background} after the necessary background.

Our technical contribution is to exploit these connections to prove several new results improving existing bounds as well as finding new applications for Mnets, 
$\e$-containers and $\e$-brackets. These include improved bounds on the size of $\e$-brackets and $\e$-containers with optimal dependence on the ambient dimension 
and showing the existence of $\lambda$-heavy $\e$-Mnets for arbitrary $\lambda$. We proceed to give several applications of our improved bounds, such as in  
distributed learning of halfspaces and distributed linear programming, and the smoothed analysis of online and differentially private learning. We also extend the 
results of~\cite{MR14,DuttaGJM19}, who showed the existence of $\Lambda$-heavy Mnets for a fixed $\Lambda\in (0,1/2)$, to show the existence of $\lambda$-heavy Mnets for any given $\lambda\in (0,1)$. These results follow from the new connections between brackets, containers and Mnets we have developed in this paper.



Our general bounds are in terms of shallow cell complexity and Property $\M$ and so can seem somewhat abstract. 
Therefore we are deferring the conceptual connections in their full generality to Section~\ref{sec:background}. 
For a set system $(X,\calR)$, its \emph{projection} on to a subset $Y\subset X$ of the ground set is the 
system $(Y,\calR_{\mid Y})$, where      
    $\calR_{\mid Y} := \left\{ R \cap Y \; \mid \; R \in \calR \right\}$. The \emph{VC dimension} of $(X,\calR)$ is 
the size of the largest subset $Y\subset X$, such that $\calR_{\mid Y} \equiv 2^Y$, i.e. the entire power set of $Y$ is 
expressible as a collection of intersections with members of the family $\calR$.
In this section, we will present a more simplified version of the structural results,
in terms of the VC dimension, and give applications of these results to online and distributed learning. 


\subsection{Bounds for Semi-algebraic Set Systems}
        
        Semi-algebraic set systems, see the Definition~\ref{def-semi-algebraic-set-systems}, are systems where the family of subsets can be described as the intersection of the ground set with a semi-algebraic 
        family of inequalities, that is, inequalities which can be formulated using a constant number of Boolean operations between 
        polynomials of bounded degree. These include halfspaces, balls, axis-parallel boxes, $k$-polytopes (where $k$ is a constant), etc. 
        A set system $(X, \calR)$ has \emph{shallow-cell complexity}
        $\psi(\cdot, \cdot)$ if for any finite subset $Y$ of $X$, the number of subsets of $Y$ of size at most $\ell$, with $\ell \leq |Y|$, that can arise as intersections with $\calR$ 
        is at most $ |Y| \cdot \psi(|Y|, \ell)$.
        We refer the reader to Section~\ref{sec:background} for 
        more precise definitions and some further examples.

        For semi-algebraic set systems of bounded shallow cell complexity, our bounds for Mnets, containers and brackets can be stated more explicitly, 
        as given below. 

\paragraph{Mnets of arbitrary heaviness.}
        The Mnet construction of Mustafa and Ray~\cite{MR14} as well as those obtained in~\cite{DuttaGJM19} are $\lambda$-heavy where 
        $\lambda\leq 1/2$. In fact in the case of the 
        Mnets obtained in~\cite{DuttaGJM19} 
        $\lambda$ is given by the multilevel polynomial partitioning theorem, and depends inversely 
        polynomially on the ambient dimension, the maximum degree of the polynomial family, and the number of allowed Boolean operations. 
        A natural question that arises is, can the heaviness of the constructed Mnets be improved beyond $1/2$ or even be made arbitrarily close 
        to $1$? A priori, this does not seem possible using the previous techniques, 
        as these rely on an application of the pigeonhole principle to choose a region from an integral number of regions, all of which are enclosed by a range. 
        Thus, in the best case, there are 2 regions inside a range and one is chosen, which gives $\lambda=1/2$.

%
%
%
%
        Our first result is that for semi-algebraic systems, Mnets of arbitrarily small heaviness 
        can be boosted to get Mnets of any desired heaviness $\lambda$. 
        This extends and generalizes the results of~\cite{MR14} and~\cite{DuttaGJM19}, whose techniques, as we observed earlier, cannot 
        give Mnets of heaviness more than $1/2$.
        
        \begin{theorem}
        [Informal statement: Mnet for semi-algebraic set system]
        \label{thm:inf-semi-algebraic-lambda-mnets}
        Let $X \subset \R^{d}$, and 
        $(X, \calR)$ be a set system induced by semi-algebraic regions in $\R^{d}$ of constant complexity with VC dimension $d_{0}$. 
         Then there exists $\lambda$-heavy $\eta$-Mnets $\calM$ of $(X, \calR)$ of size at most  
         \[|\calM| \leq \pth{\frac{2}{1-\lambda}}^{c_{1}d_{0}}\times \pth{\frac{c_{2}}{\eta}}^{2d_{0}},\]
         where $c_{1}$ depends only on $d$ and $c_{2}$ is an absolute constant. 
         (For a more precise bound in terms of the shallow cell complexity 
         see Theorem~\ref{thm:semi-algebraic-lambda-mnets})
\end{theorem}


         
\paragraph{Containers.} 
        Generalizing the results of Braverman et al.~\cite{BravermanKMS21} showing the existence of containers for points and halfspaces, we show that 
        containers can be obtained for semi-algebraic set systems. 
        
        \begin{theorem}[Informal statement: Containers for semi-algebraic set systems]\label{thm:inf-eps-container-semialg-bd}
            Let $X \subset \R^{d}$, and $(X, \calR)$ be a set system induced by semi-algebraic regions in $\R^{d}$ of constant complexity with VC dimension $d_{0}$. Then there exists an $\e$-container $\calC$ for $(X, \calR)$ of size at most
            \[
                |\calC| \leq \pth{\frac{2}{\e}}^{cd_{0}},
            \]
            where $c$ depends only on $d$. (For a more precise bound in terms of the shallow cell complexity see Theorem~\ref{thm:eps-container-semialg-bd})
         \end{theorem}

        While the bounds on containers for points and halfspaces in~\cite{BravermanKMS21} can be shown to hold for semi-algebraic systems using operations like Veronese mappings and lifts,  
        such operations can blow up the ambient dimensionality -- which appears in the exponent in the bounds -- by a polynomial factor. 
        The general version of Theorem~\ref{thm:eps-container-semialg-bd} (see Theorem~\ref{thm:eps-container-semialg-bd}) gives direct bounds on the size of the container family in terms of shallow cell complexity, which in some case has a lower dimensionality, and 
        therefore better bounds, than those of~\cite{BravermanKMS21}. This is usually the case for things like $\e$-nets as shallow cell complexity captures the combinatorial complexity of set systems at a much finer scale than for say VC dimension~\cite{AronovES09,Varadarajan10,ChanGKS12,MDG17,MustafaV17}.

        More specifically, for set system of points and halfspaces in $\R^{d}$ we obtain the following improved bound for containers for points and halfspaces. 
         
        \begin{theorem}[Improved container bounds for points and halfspaces]\label{thm:eps-container-points-halffspaces-Rd-bd}
            Let $X \subset \R^{d}$, and $\e \in (0,1)$. Then there exists 
            a collection of subset $\calC$ of $X$ of size at most $\left( \frac{2}{\e}\right)^{O(d)}$ such that for all halfspaces $h$ of $\R^{d}$ there exists $C_{h} \in \calC$ such that 
            \[
                X\cap h \subseteq C_{h}\;\;\mbox{and}\;\; |C_{h}\setminus \left( X\cap h\right)| \leq \e |X|. 
            \]
         \end{theorem}
        
        The above theorem removes the multiplicative factor of $d^{O(d)}$ which appears in the bounds of Braverman et al.~\cite{BravermanKMS21}, thus 
        significantly improving the dependence on the ambient dimension, from superexponential to exponential. It is easier to see the improvement in Theorem~\ref{thm:eps-container-points-halffspaces-Rd-bd} if we fix some $\e \in (0,1)$ and make $d$ tend to infinity. This dynamic plays a crucial role in getting the optimal communication complexity of distributed learning of halfspace problem, see Theorem~\ref{thm:1}.

        

\paragraph{Uniform brackets.}
Finally, we combine Theorems~\ref{thm:inf-semi-algebraic-lambda-mnets} and~\ref{thm:inf-eps-container-semialg-bd} to get explicit bounds on the size of $\e$-brackets.
        \begin{corollary}[Informal statement: Bracket for semi-algebraic set systems]
         \label{cor:bracket-bd}
         Let $X \subset \R^{d}$, and $(X, \calR)$ be a set system induced by semi-algebraic regions in $\R^{d}$ of constant complexity with VC dimension $d_{0}$. 
         Then there exists an $\e$-bracket $\calB$ for $(X, \calR)$ of size at most
            \[
                |\calC| \leq \pth{\frac{2}{\e}}^{cd_{0}},
            \]
            where $c$ depends only on $d$. (For a more precise bound in terms of the shallow cell complexity see Theorem~\ref{thm:bracket-bd})
         \end{corollary}
         
    
    It is a simple exercise to see that any $\e/2$-container for points and halfspaces in $\R^{d}$ is also an $\e$-bracket for the same set of points and halfspaces in $\R^{d}$. Therefore, we get the following result directly from Theorem~\ref{thm:eps-container-points-halffspaces-Rd-bd}.
    
    \begin{corollary}[Improved bracketing bounds for points and halfspaces]\label{thm-points-halfspaces-brackets}
        Let $X \subset \R^{d}$, and $\e \in (0,1)$. Then there exists 
        a collection of subset $\calB$ of $X$ of size at most $\left( \frac{2}{\e}\right)^{O(d)}$ such that for all halfspaces $h$ of $\R^{d}$, there exist sets $B^{-}_{h}$ and $B^{+}_{h}$ in $\calB$ such that 
        \[
            B^{-}_{h} \subseteq X\cap h \subseteq B^{+}_{h}\;\;\mbox{and}\;\; |B^{+}_{h}\setminus B^{-}_{h}| \leq \e |X|. 
        \]
    \end{corollary} 


The above theorem directly implies an improved {\em distribution-free}\footnote{Our upper bound on the $\e$-bracketing number for halfspaces in $\R^{d}$ is called distribution-free because the bound does not depend on the probability measure $\mu$.} bound for any collection of halfspaces in $\R^{d}$. See, Braverman et al.~\cite{BravermanKMS21} and Haghtalab et al.~\cite{HaghtalabRS20}.

\begin{corollary}[Improved bracketing number for halfspaces]
    \label{theorem-improved-bracketing-number-halfspaces-measure}
    Let $\mathcal{H}$ be a family of halfspaces in $\R^{d}$. For all $\e \in (0,1)$, and probability measure $\mu$ over $\R^{d}$ we have 
    $\brac(\calH,\mu,\e) \leq \left( 2/\e\right)^{O(d)}$.
\end{corollary}

Braverman et al.~\cite{BravermanKMS21} and Haghtalab et al.~\cite{HaghtalabRS20} showed that distribution-free $\e$-bracketing number for halfspaces in $\R^{d}$ is $\left( \frac{d}{\e}\right)^{O(d)}$. Note that our result is an improvement over this bound by a factor of $ d^{O(d)}$. More detailed calculations reveal the constant in the $O(d)$-exponent to be less than $7.03$ in our case.
    
Further, the following lower bounds show that the upper bounds established above are best possible up to dimension-independent constants in the exponent
\begin{theorem}[Lower bounds for $\e$-containers] \label{thm:lb-eps-cont}
    There exists $C_{d}$ that depends only on $d$ such that the following holds:
    \begin{itemize}
        \item Given positive integers $d \geq 2$, $n$, and $\e \in (0,1)$, there exists a set $X$ of $n$ points in $\mathbb{R}^{d}$ such that any $\e$-container for the set system induced by the set $X$ and halfspaces has size at least 
        $$
            C_{d}\cdot \frac{1}{\e^{\left\lfloor (d+1)/3 \right\rfloor}}.
        $$
        \item
            For all integers $d \geq 2$, $n \geq 0$, and $\e \in (0,1)$, there exists a set $Y$ of $n$ points in $\mathbb{R}^{d}$ such that any $\e$-container for the set system induced by the set $Y$ and hyperplanes has size at least 
        $$
            C_{d}\cdot \frac{1}{\e^{d}}.
        $$
    \end{itemize}
%
\end{theorem}

\begin{remark}
    \begin{enumerate}
        \item
            The lower bounds in Theorem~\ref{thm:lb-eps-cont} directly follow from \cite[Corollary~4.1]{MR14} and \cite[Theorem~4.6]{DuttaGJM19}.
            
        \item
            Note that the set systems induced by halfspaces and hyperplanes in $\R^{d}$ have VC dimension  $d+1$ and $d$ respectively.
        
        \item
            Family of hyperplanes and halfspaces in $\R^{d}$ belong to the semi-algebraic family $\Gamma_{d,1,1}$.
    \end{enumerate}
\end{remark}

\subsection{Applications}

Our improved bounds have applications in several areas such as the smoothed analysis of online learning algorithms as well as 
in distributed learning algorithms, e.g. the disjointedness of convex bodies and LP feasibility. Some of these applications are described below.


\paragraph*{Distributed learning of halfspaces.} 
Linear classifiers are objects of central importance in many machine learning algorithms, beginning from the original \emph{perceptron} model of 
Rosenblatt~\cite{Roosenblatt58} to modern algorithms like neural networks, kernel machines, etc. A basic problem in machine learning therefore, 
relates to the learning of linear classifiers, which are essentially halfspaces. The {distributed learning of halfspaces} problem has received 
considerable attention~\cite{ChangZWBLQC07, ForeroCG10, McDonaldHM10, BalcanBFM12, DaumePSV12, KaneLMY19, BravermanKMS21}. Balcan et al.~\cite{BalcanBFM12} 
and Daum{\'{e}} III et al.~\cite{DaumePSV12} proved an upper bound of  $O(d\log^{2}n)$ bits on the communication complexity of learning halfspaces over a domain of $n$ points in $\mathbb{R}^d$, and Kane et al.~\cite{KaneLMY19} proved that 
any randomized protocol for the above problem will require $\Omega\left( d + \log n\right)$ bits of communication. Braverman et al.~\cite{BravermanKMS21} gave 
an improved a deterministic protocol with communication complexity $O(d\log d\log n)$, and proved an almost matching lower bound of $\Omega(d\log(n/d))$.

Let $U$ be a known set of $n$ points in $\mathbb{R}^{d}$. In {\em distributed learning of halfspaces} problem, two players, Alice and Bob are given sets $S_{a}$ and 
$S_{b}$ where $S_{a}, \, S_{b} \subseteq U \times \{\pm 1\}$ respectively such that the sets 
$\left\{(x, +1) \in S_{a} \cup S_{b} \; : \; x \in U \right\}$ and 
$\left\{ (x, -1) \in S_{a} \cup S_{b} \; : \; x \in U \right\}$ can be separated 
by a hyperplane in $\mathbb{R}^{d}$. The goal is for both the players, using  to agree classifier $H:U \to \{\pm 1\}$, such that
\begin{itemize}
\item
if $(x, +1) \in S_{a} \cup S_{b}$ then $H(x) = +1$, and 

\item
    if $(x, -1) \in S_{a} \cup S_{b}$ then $H(x) = -1$.
\end{itemize}
%
%

Using the communication protocol of Braverman et al.~\cite{BravermanKMS21} for the problem
together with Corollary~\ref{thm:eps-container-bd} we get the following upper bound which tightly\footnote{That is, up to a universal multiplicative constant.} meets the lower bound when $n\geq d^{1+\Omega(1)}$.

\begin{theorem}
\label{thm:1}
Let $U$ be a known $n$-sized subset of $\mathbb{R}^{d}$. Then, there exists a deterministic protocol for Learning Halfspaces over $U$ with
communication complexity $O\left(d \log n\right)$ bits.
\end{theorem}
We note that in this context previous works typically assume that the number of domain points $n$ is much larger than the euclidean-dimension, 
and often even that $n=\exp(d)$.
(Consider e.g.\ the natural case when the domain $$U=\{0,1\}^d$$ consists of all binary vectors in $\mathbb{R}^d$.)
In such cases, the above upper bound completely resolves the communication complexity of distributed learning of halfspaces.

\paragraph*{Distributed convex set disjointness problem and LP feasibility.}
Kane et al.~\cite{KaneLMY19} introduced the {\em distributed convex set disjointness} problem in communication complexity, where, like in the 
case of distributed learning of halfspaces, there is a known $n$-sized domain $U \subset \R^{d}$ and 
two parties Alice and Bob are given as inputs $S_{a}$ and $S_{b}$, with $S_{a}, \, S_{b} \subset U$, respectively. The goal is for both parties 
to decide if the convex hulls\footnote{For any subset $S$ of $\R^{d}$, {\em convex hull} of $S$ will be denoted by $\mathrm{conv}(S)$.} 
$\mathrm{conv}(S_{a})$ and $\mathrm{conv}(S_{b})$ intersect or not. 
Note that the distributed convex set disjointness problem is equivalent to the fundamental problem of 
two-party {\em distributed Linear Programming (LP) feasibility}. For a more detailed discussion on this equivalence, see~\cite{BravermanKMS21}. 

Vempala et al.~\cite{VempalaWW20} gave the first $O\left(d^{3}\log^{2} n\right)$ upper bound for the distributed convex set 
disjointness problem, and $\Omega\left(d \log n\right)$ and $\Omega\left( \log n \right)$ lower bounds for the 
deterministic and randomized settings respectively.
Braverman et al.~\cite{BravermanKMS21} gave an improved $O\left( d^{2} \log d \log n\right)$ upper bound for the distributed 
convex set disjointness problem, and they also proved a randomized $\Omega(d\log n)$ bits lower bound. Observe that 
Theorem~\ref{thm:2} gives an $\log d$ factor improvement over the bound of Braverman et al.~\cite{BravermanKMS21}.

Using the Braverman et al.~\cite{BravermanKMS21} communication protocol for the distributed convex set disjointness problem 
together with Corollary~\ref{thm-points-halfspaces-brackets} we get the following result.
\begin{theorem}
\label{thm:2}
Let $U$ be an $n$-sized subset of $\mathbb{R}^{d}$. Then there exists a deterministic communication protocol for Convex 
Set Disjointness problem over $U$ with communication complexity $O\left(d^{2} \log n\right)$ bits.
\end{theorem}

\paragraph{Improved bracketing number and online algorithms.}
The {\em bracketing number} of a set system is a fundamental tool in statistics for proving uniform laws of large numbers for 
empirical processes~\cite{adamsnobel2010}. More recently, 
Haghtalab, Roughgarden and Shetty~\cite{HaghtalabRS20} used bracketing numbers for smoothed analysis of online and differentially private learning algorithms. 



Haghtalab, Roughgarden and Shetty~\cite{HaghtalabRS20}, using the Braverman et al.~\cite{BravermanKMS21} $\e$-container bound for points and halfspaces, showed that 
 \begin{eqnarray}
        {N}_{[\;]}\left( \mathcal{P}^{n,d}, \mu, \epsilon\right) \leq \exp\left( c_{1} n^{d}\ln \left( n^{d}/\epsilon \right)\right), \; \mbox{and}\;
        {N}_{[\;]}\left( \mathcal{Q}^{d, k}, \mu, \epsilon\right) \leq \exp\left( c_{2} nk\ln \left( nk/\epsilon \right)\right).
\end{eqnarray}
where
$\mathcal{P}^{n,d}$ denotes the class of {\em $d$-degree polynomial threshold functions} in $\R^{n}$ and $\mathcal{Q}^{n, k}$ be the class of 
{\em $k$-polytopes} in $\mathbb{R}^{n}$, and $c_{1}$ and $c_{2}$ are absolute constants. Using Corollary~\ref{theorem-improved-bracketing-number-halfspaces-measure}, together with \cite[Theorem~3.7]{HaghtalabRS20}, we can directly improve the distribution-free bounds 
for $\e$-bracketing numbers:
\begin{theorem}
\label{thm-improved-bound-threshold-func-k-polytopes}
    Let $(\R^{n}, \Omega, \mu)$ be a probability space. Then
    \begin{enumerate}
        \item 
            ${N}_{[\;]}\left( \mathcal{P}^{n,d}, \mu, \epsilon\right) \leq \exp\left( c_{1} n^{d}\ln \left( 1/\epsilon \right)\right)$, where $c_{1}$ is an absolute constant.
        
        \item
            ${N}_{[\;]}\left( \mathcal{Q}^{d, k}, \mu, \epsilon\right) \leq \exp\left( c_{2} nk\ln \left( 1/\epsilon \right)\right)$, where $c_{2}$ is an absolute constant.
    \end{enumerate}
\end{theorem}

   The notion of \emph{regret minimization} is a standard measure of the effectiveness of machine learning algorithms. In the context of 
   \emph{online learning}, which has arisen from the need to design learning algorithms robust to small changes in the input data, worst-case 
   online learnability is characterized by having finite Littlestone dimension~\cite{AlonLMM19,Ben-DavidPS09,BunLM20}. However, this can be a very restrictive condition, 
   as there are instances of problems which have constant VC dimension, yet their Littlestone dimension is infinite~\cite{DBLP:conf/compgeom/AlonHW87,AlonLMM19,Ben-DavidPS09,BunLM20}. 
   Recently, 
   going beyond worst-case analysis, Haghtalab et al.~\cite{HaghtalabRS20} have introduced the \emph{smoothed analysis} paradigm of Spielman-Teng~\cite{SpielmanT04} 
   to the context of online learning algorithms. Using this paradigm, they designed online learning 
   no-regret algorithms for several problems with infinite Littlestone dimension, even for the case of \emph{adaptive adversaries}, provided 
   the adversaries choose from a $\sigma$-smooth distribution. For an introduction to online regret minimization against an $\sigma$-smoothed adversary see~\cite{HaghtalabRS20}. 
   
   Using Theorem~\ref{thm-improved-bound-threshold-func-k-polytopes}, together with  \cite[Theorem~3.3]{HaghtalabRS20}
   we will get the following improved online algorithm whose regret against an adaptive $\sigma$-smoothed adversary on $\mathcal{P}^{n,d}$ and 
   $\mathcal{Q}^{n, k}$ satisfies:
\begin{theorem}\label{thm-improved-regret bound}
    There exists an online algorithm against an adaptive $\sigma$-smoothed adversary whose regret after $T$-steps is
    \begin{enumerate}
        \item 
            $O\left(\sqrt{T\cdot \mathrm{VCdim}\left( \mathcal{P}^{n,d}\right)\log \frac{T}{\sigma}} \right)$ if the class of functions is $\mathcal{P}^{n,d}$.

        \item
            $O\left(\sqrt{T\cdot \mathrm{VCdim}\left( \mathcal{Q}^{n,k}\right)\log \frac{T}{\sigma}} \right)$ if the class of functions is $\mathcal{Q}^{n,d}$.
    \end{enumerate}
\end{theorem}

\begin{remark}
    Theorem~\ref{thm-improved-regret bound} is an improvement over \cite[Corollary~3.8]{HaghtalabRS20} where the regret bounds were 
    \[
        O\left(\sqrt{T\cdot \mathrm{VCdim}\left( \mathcal{P}^{n,d}\right)\left(\log \frac{T}{\sigma} + \log \mathrm{VCdim}\left( \mathcal{P}^{n,d}\right)\right)} \right)
    \]
    and 
    \[
        O\left(\sqrt{T\cdot \mathrm{VCdim}\left( \mathcal{P}^{n,d}\right)\left(\log \frac{T}{\sigma} + \log \mathrm{VCdim}\left( \mathcal{P}^{n,d}\right)\right)} \right)
    \]
    for the class of functions $\mathcal{P}^{n,d}$ and $\mathcal{Q}^{n,d}$ respectively. 
\end{remark}

\section{Notations, definitions, and background results}
\label{sec:background}

In this section, we formally define various notions, definitions, and necessary background used in this work.

\subsection*{Notations}

We use the following notational conventions throughout the paper. The complement of a set $R$ with respect to some ground set $X$, 
is denoted by $R^{c} := X\setminus R$. The \emph{complement family} of a family of subsets $\calR$ of the ground set $X$, is 
denoted by $\calR^{(c)} := \left\{ X\setminus R \; \mid\; R \in \calR \right\}$. 
Given two sets $A$ and $B$, $A\Delta B$ denotes the {\em symmetric difference} between the two sets, that is, $A\Delta B = (A\setminus B) \cup (B\setminus A)$.
For a family of subsets $\calR \subset 2^X$, 
the sub-family of subsets of size at most $t$, is denoted by 
        $\calR^{\leq t} := \{R \; \mid \; R \in \calR \;\mbox{and}\; |R| \leq t\}$, and similarly we can define $\calR^{\geq t}$, $\calR^{<t}$ and $\calR^{>t}$.
For an open interval $I= (a,b)$, the family of ranges $R\in \calR$ with $|R|\in I$, are denoted by $\calR^{(a,b)}$. 
We use the corresponding notations for closed and half-open intervals.




\subsection*{Definitions and background results}

        \begin{lemma}[Sauer-Shelah Lemma~\cite{DBLP:journals/jct/Sauer72,S-Shelah72}]
        \label{l:sau-she-71}
            Let $(X,\calS)$ be a set system with $|X|=n$, having VC dimension $d_{0}$. Then the number of sets in the family $\calS$ satisfies
            \begin{equation*}
                |\calS| \leq \sum_{i=0}^{d_{0}} {n\choose i} \leq \left(\frac{en}{d_{0}}\right)^{d_{0}}.
            \end{equation*}
        \end{lemma}

As mentioned in the Introduction, 
        the \emph{shallow cell complexity} is a finer characterization of the complexity of a set system than its VC dimension, and it has been shown that for 
        most geometric set systems, the shallow cell complexity yields optimal bounds on the sizes of $\e$-nets and related structures. 
        \begin{definition}
        \label{def:shall-cell-complx}
         A set system $(X, \calR)$ has \emph{shallow-cell complexity}
         $\psi(\cdot, \cdot)$ if for any finite subset $Y$ of $X$, we have
         that the number of subsets in $\calR|_Y$ of size at most $\ell$  
         is at most $ |Y| \cdot \psi(|Y|, \ell)$.
        \end{definition}
        Next, we formally define semi-algebraic families in $\R^d$ and the set systems generated by them.

        \begin{definition}\label{def-semi-algebraic-families}
         Given $d,s,\Delta\in \N$, the \emph{semi-algebraic family} $\Gamma_{d,\Delta,s}$ denotes the class of all subsets of $\R^d$ which can be defined 
         by a boolean formula with at most $s$ Boolean operations (i.e.\ union, intersection, and complementation) on sets definable by polynomial inequalities of the type $f(x) \geq 0$, where $f:\R^d\to \R$ is a $d$-variate 
         polynomial of degree at most $\Delta$.
        \end{definition}
	
        \begin{definition}\label{def-semi-algebraic-set-systems}
            A range space $(X,\calR)$ where $X$ is a set of $n$ points in $\R^d$, is said to be a \emph{semi-algebraic system} generated by $\Gamma_{d,\Delta,s}$
         if for every $R\in\calR$, there exists a set $S\in \Gamma_{d,\Delta,s}$ such that $R=S\cap X$.
        \end{definition}


        The existence of Mnets of bounded size for semi-algebraic systems of bounded VC dimension was proved in~\cite{DuttaGJM19}.  
        \begin{theorem}[Mnets for semialgebraic set systems~\cite{DuttaGJM19}]\label{thm:mnets-semialgebraic-set-systems}
            Let $d$, $d_{0}$, $\Delta$ and $s$ be integers and $(X, \calR)$ be a semialgebraic set system generated by $\Gamma_{d,\Delta,s}$, with $|X| = n$ 
        and VC dimension at most $d_{0}$. Then there exists a constant $\Lambda=\Lambda_{d,\Delta,s} \in (0, 1)$ such that for any $\e> 0$ the system $(X, \calR)$ has a 
        $\Lambda$-heavy $\e$-Mnet of size
            \[
                K\cdot \left(\frac{c}{\e} \right)^{d_{0}},
            \]
        where $K=K_{d,\Delta, s}$ depends only $d$, $\Delta$ and $s$, and $c$ is an absolute constant independent of $n,d,d_0,s$ and $\Delta$.
        \end{theorem}


We shall require the following lower bound for Mnets, proved by Mustafa and Ray~\cite{MR14}.
\begin{theorem}[Mnets lower bound for points and halfspaces~\cite{MR14}]
\label{thm:lb-mr-pthfsp}
    Given integers $d\geq 2$ and $n\geq 0$, there exists a set of $n$ points in $\R^d$, such that for any $\e\in (0,1)$ the set system generated by half-spaces cannot have 
    Mnets of size less than 
    $$
        C_{d}\cdot\frac{1}{\e^{\lfloor (d+1)/3\rfloor}}
    $$
    where $C_{d}$ depends only on $d$.
\end{theorem}

The following upper bound was also proved in~\cite{MR14}.
\begin{theorem}[Mnets for points and halfspaces~\cite{MR14}] \label{thm:mustafa-ray-14}
    Let $X$ be a set of $n$ points in $\R^{d}$, and $(X, \calR)$ be a primal set system generated by halfspaces in $\R^{d}$. 
Then there exists an $\frac{1}{2}$-heavy $\e$-Mnet for $\calR$ of size at most $\left( \frac{O(1)}{\e}\right)^{\lfloor d/2\rfloor}$.
\end{theorem}

Next, we come to packing bounds for set systems having bounded VC dimension. 

\begin{definition}[Shallow Packings]
\label{defn:shall-pack}
    Let $(X, \calR)$ be a set system, and $\delta$ and $k$ be positive integers. 
    \begin{itemize}
        \item {\bf $\delta$-packing:} A subset of ranges $\calP \subseteq \calR$ is a $\delta$-packing if for any two distinct sets $R_{1}$ and $R_{2}$ in 
        $\calP$ we have $|R_{1}\Delta R_{2}| > \delta$.
        
        \item {\bf $k$-shallow $\delta$-packing:} $\calP \subseteq \calR$ is a $k$-shallow $\delta$-packing if $\calP$ is a $\delta$-packing, and for 
        all $R \in \calP$ we have $|R| \leq k$.
    \end{itemize}
\end{definition}

Haussler~\cite{Haussler92spherepacking} proved the following seminal result about packing and VC dimension.
\begin{theorem}[Haussler's Packing Lemma~\cite{Haussler92spherepacking}]\label{thm:haussler-packing-lemma}
    Let $(X, \calR)$ be a set system with $|X| = n$ and VC dimension at most $d_{0}$ and let $\delta \leq n$.
Then, if $\calS \subseteq \calR$ is a $\delta$-packing then $|\calS| \leq \left( \frac{cn}{\delta}\right)^{d_{0}}$, where $c$ is an absolute constant. 
\end{theorem}

Recently, following several developments, Mustafa~\cite{M-shallow-packing-2016} gave optimal packing bounds for 
set systems in terms of their shallow cell complexity.
\begin{theorem}[Shallow Packing Lemma~\cite{M-shallow-packing-2016}]
\label{thm:shallow-packing-lemma}
    Let $(X, \calR)$ be a set system with $|X| = n$ and shallow cell complexity  $\varphi_{\calR}$. If 
VC dimension of $(X, \calR)$ is at most $d_{0}$, and $(X, \calR)$ is a $k$-shallow $\delta$-packing then 
    \[ 
        \frac{24d_{0}n}{\delta} \cdot \varphi_{\calR}\left( \frac{4d_{0}n}{\delta}, \frac{12d_{0}k}{\delta}\right).
    \]
\end{theorem}

A $\delta$-packing $\calP \subseteq \calR$ is \emph{maximal} if no other range in $\calR$ can be added to $\calP$ so 
that the resulting family is still a $\delta$-packing. Maximal $\delta$-packings have the following property, which shall be very useful for us:
\begin{proposition}
\label{propn:nn-max-pack}
Let $(X, \calR)$ be a set system, and $\calP\subset \calR$ be a maximal $\delta$-packing.
For every range $R\in \calR$, there exists a range $P\in \calP$, called a \emph{nearest neighbour} of $R$ in $\calP$, such that $|R\Delta P|\leq \delta$.
\end{proposition}
\begin{proof}
   The proof follows from the inclusion-maximality of $\calP$. Given $R\in \calR$, if $R\in \calP$, then we are done, as $|R\Delta R|=0 \leq \delta$, and therefore $R$ is 
its own nearest neighbour in $\calP$. Suppose for every range $P\in \calP$ we have $|R\Delta P|> \delta$. Then $R$
could be added to $\calP$ to get a larger $\delta$-packing, which would contradict the maximality of $\calP$.
\end{proof}

\section{General Theorems}

Now we come to our general results -- the conceptual connections between brackets, containers and Mnets.
To state our results in their full generality, we first need to define \emph{Property $\M$}.

\subsection{Property $\M$: Brackets, Containers, and Mnets}

Let $(X,\calR)$ be a set-system with a finite VC dimension $d_0$.
We say that $(X,\calR)$ satisfies \emph{Property $\M$} with bound $f(.)$, if
there exists 
$\Lambda\in (0,1)$ and a function $f=f_{\Lambda,d_0}:(0,1)\to \R_+$, such that for any finite subset $Y$ of $X$ and for all $\e \in (0,1)$, the set system
$(Y,\calR_{\mid Y})$ has a $\Lambda$-heavy $\e$-Mnet of size at most $f(\e)$.
The key distinction here from the definition of Mnets is that $\Lambda$ is fixed, whereas in the definition of Mnets
we require $\Lambda$-heavy $\e$-Mnets of bounded size for \emph{every} $\Lambda$ and every $\e$.


\begin{remark}It is not too hard to see that having Property $\M$ is stronger than having bounded VC dimension, since having Mnets of bounded size implies having $\e$-nets 
of bounded size (see~\cite{DuttaGJM19} for an optimal extraction of $\e$-nets from Mnets), which implies bounded VC dimension. However, as shown 
in~\cite{DuttaGJM19}, all geometrically defined set systems having bounded semi-algebraic complexity, have this property. Thus Property $\M$ lies 
somewhere between having bounded VC dimension and having bounded semi-algebraic complexity, and with this new definition we can get a VC-type theory connecting Mnets, containers and brackets. It might be interesting to determine whether property $\M$ can be characterized with a simple combinatorial parameter, in the spirit of the VC dimension.
\end{remark}

We now state our general results. 
Our first result is a \emph{boosting} lemma showing that the existence of a $1/2$-Mnet in a set system
implies the existence of an $\e$-Mnet for any $\e\in (0,1)$.

    \begin{lemma}[$\e$-Boosting Lemma]
    \label{lemma:mnets-1/2-epsilon-Mnet}
        Let $(X,\calR)$ be a set system with VC dimension at most $d$ such that for all $Y \subseteq X$, the set system 
        $(Y, \calR\mid_{Y})$ has a $\lambda$-heavy $\frac{1}{2}$-Mnet of size at most $\Delta$, for some $\lambda\in (0,1)$. 
        Then for all $\epsilon \in (0,1)$ and $\eta \in (0,1)$, 
        $(X, \calR)$ has an $\lambda'$-heavy $\epsilon$-Mnet of size at most 
        $O\left( \frac{(2c)^{d}\Delta}{\epsilon^{d}}\max\left\{2^{d}, \frac{1}{\eta^{d}}\right\} \right)$ where $\lambda' = \lambda (1 - \eta)$. 
    \end{lemma}
    
    Next, in Lemma~\ref{l:mnet-implies-container}, we show that Mnets and containers have a complementary relationship 
    -- the existence of Mnets in a set system
    implies the existence of containers in the complementary system, and vice versa.
    
    \begin{lemma} 
        \label{l:mnet-implies-container} 
        Given a set system $(X,\calR)$ with $|X| = n$, $\delta_0 \in (0,1]$ and $\lambda \in (0,1)$. 
        \begin{enumerate}[{\bf (a)}]
            \item 
                If $\calM$ is a $\lambda$-heavy $(1-\delta_{0})$-Mnet for $\left( X, \left( \calR^{\leq \delta_0 n} \right)^{(c)} \right)$ 
                then $\calM^{(c)}$ is a $(1-\lambda+\lambda\delta_{0})$-container family for $\left(X, \calR^{\leq \delta_{0}n}\right)$. 
            
            \item
                If $\calC$ is a $(1-\lambda)$-container family for $\left( X,\calR^{\leq\delta_0 n} \right)$, 
                then $\calC^{(c)}$ is an $\left( \lambda - \delta_{0}\right)$-heavy $\left( 1 - \delta_{0} \right)$-Mnet 
                for $\left(X, \left( \calR^{\leq \delta_{0} n}\right)^{(c)} \right)$.
        \end{enumerate}
    \end{lemma}


Our next result can be thought of as an analogue of Lemma~\ref{lemma:mnets-1/2-epsilon-Mnet} for the heaviness, 
that is, a $\Lambda$-heavy $\e$-Mnet for some $\Lambda\in (0,1)$ can be boosted to $\lambda$-heavy $\e$-Mnets for any $\lambda \in (0,1)$.
Given $k$, $l\geq 0$, let $p_{\calR}(k,l)$ be the maximum size of 
an {\em $l$-shallow $k$-packing} of $\calR$ (see Definition~\ref{defn:shall-pack}).

    \begin{theorem}[Arbitrarily heavy Mnets]
    \label{thm:any-lambda-mnets}
        Let $(X,\calR)$ be a set system with $|X|=n$, having VC dimension at most $d_0$, and Property $\M$ 
        for some $\Lambda\in (0,1)$, with bound $M^*(.)=M^*_{\Lambda,d_0}(.)$. 
        Then given any $\eta,\lambda\in (0,1)$, there exists $t_0=t_0(\lambda,\Lambda) := 1+\frac{\log 4/(1-\lambda)}{\log(1-\Lambda/2)^{-1}}$, and 
        sequences $(\delta_k)_k$, $(l_k)_k$ $k=0,1,\ldots$, with $0\leq \delta_k,l_k\leq 1$, given by 
        $\delta_k = \pth{1+\frac{1-\lambda}{4}}^k\eta^2 $, and $l_k= \pth{1+\frac{1-\lambda}{4}}^{k+1}\eta $, 
         such that $(X,\calR)$ has a $\lambda$-heavy $\eta$-Mnet $\calM$ of size at most 
         \[
            |\calM| = M^*(1/2)^{t_0}\cdot\pth{\sum_{k\geq 0} p_{\calR}(\delta_k n,l_k n)}.
         \]
         Using Theorem~\ref{thm:haussler-packing-lemma}, the above bound implies the existence of $\lambda$-heavy $\eta$-Mnet $\calM$ of size at most
         \[
            |\calM| = \frac{4\, M^*(1/2)^{t_0}}{1-\lambda}\times \left(\frac{c}{\eta^{2}} \right)^{d},
         \]
         where the constant ``$c$'' the same as the one in Theorem~\ref{thm:haussler-packing-lemma}.
         
        \end{theorem} 
        
Following this, we show that Property $\M$ in a set system, i.e. having Mnets of bounded size, 
also implies the existence of $\e$-containers in the complement system. This quantifies the complementary 
relation proved in Lemma~\ref{l:mnet-implies-container}.
        \begin{theorem}[$\e$-Containers]
        \label{thm:eps-container-bd}
            Let $(X,\calR)$ be a set system with $|X|=n$, having VC dimension at most $d_0$, such that $(X,\calR^{(c)})$ has Property 
         $\M$ with bound $\bar{M}^*(.)=\bar{M}^*_{\Lambda,d_0}(.)$, for some $\Lambda\in (0,1)$.
         Then given any $\e\in (0,1)$, there exists $t_0=t_0(\e,\Lambda) := 1+\frac{\log 1/\e}{\log(1-\Lambda/2)^{-1}}$, and sequences 
         $(\delta_k)_k$, $(l_k)_k$ $k=0,1,\ldots$, 
         given by $\delta_k = \pth{1+\e}^k\e^2 $, and $l_k= \pth{1+\e}^{k+1}\e $, 
         $0<\delta_k,l_k\leq 1$, such that $(X,\calR)$ has an $\e$-container $\calC$ of size at most 
         \[|\calC| = \bar{M}^*(1/2)^{t_0}\cdot\pth{\sum_{k\geq 0} p_{\calR^{(c)}}(\delta_k n,l_k n)}.
         \]
         Using Theorem~\ref{thm:haussler-packing-lemma}, the above bound implies the existence of $\e$-container $\calC$ of size at most
         \[
            |\calC| = \frac{\bar{M}^*(1/2)^{t_0}}{\e}\times \left(\frac{c}{\e^{2}} \right)^{d},
         \]
         where the constant ``$c$'' the same as the one in Theorem~\ref{thm:haussler-packing-lemma}.
         
        \end{theorem} 
        
        The following corollary on $\e$-uniform brackets can be easily deduced by applying the above theorem on the set system and its complement set system.
        \begin{corollary}[$\e$-Uniform brackets]
        \label{cor:unif-brackets}
            Let $(X,\calR)$ be a set system with $|X|=n$, having VC dimension at most $d_0$. Additionally, assume that $(X,\calR)$ and $(X,\calR^{(c)})$ 
            both have Property $\M$ with bounds $M^*(.) = M^{*}_{\Lambda, d_{0}}(\cdot)$ and $\bar{M}^*(.) = \bar{M}^{*}_{\Lambda, d_{0}}(\cdot)$ for some $\Lambda \in (0,1)$, respectively.
         Then given any $\e\in (0,1)$, there exists $t_0=t_0(\e,\Lambda):=1+\frac{\log 2/\e}{\log(1-\Lambda/2)^{-1}}$, and sequences $(\delta_k)_k$, $(l_k)_k$ $k=0,1,\ldots$, 
         given by $\delta_k = \pth{1+\frac{\e}{2}}^k\frac{\e^2}{4}$, and $l_k= \pth{1+\frac{\e}{2}}^{k+1}\frac{\e}{2} $, 
         $0<\delta_k,l_k\leq 1$, such that $(X,\calR)$ has an $\e$-uniform bracket $\calB$ of size at most
         \[
            b(\e) := M^*(1/2)^{t_0}\cdot\pth{\sum_{k\geq 0} p_{\calR}(\delta_k n,l_k n)} + \bar{M}^*(1/2)^{t_0}\cdot\pth{\sum_{k\geq 0} p_{\calR^{(c)}}(\delta_k n,l_k n)}.
         \]
        Using Theorem~\ref{thm:haussler-packing-lemma}, the above bound implies the existence of $\e$-bracket $\calB$ of size at most
         \[
            |\calB| = \frac{M^*(1/2)^{t_0}+\bar{M}^*(1/2)^{t_0}}{\e}\times \left(\frac{c}{\e^{2}} \right)^{d},
         \]
         where the constant ``$c$'' the same as the one in Theorem~\ref{thm:haussler-packing-lemma}.
        \end{corollary} 
        
        \begin{remark}
            Note that the VC dimension of an set system $(X,\calR)$ and its complement set system $(X, \calR^{(c)})$ is same.
        \end{remark}

\subsection{Bounds for semi-algebraic set system in terms of shallow cell complexity}

The bounds on brackets, containers, and Mnets given in Section~{sec:rel-work-our-contribution} can be made more explicit using the notion of \emph{shallow cell complexity}. The corresponding theorems are 
presented below. The reader may recall the appropriate definitions from Section~\ref{sec:background}.

        \begin{theorem}
        [Mnet bounds for semi-algebraic set system]
        \label{thm:semi-algebraic-lambda-mnets}
         If $(X,\calR)$ is a semi-algebraic set system generated by $\Gamma_{d, \Delta, s}$ 
         with VC dimension $d_{0}$ and shallow cell complexity $\psi(\cdot, \cdot)$,
         then there exists $\lambda$-heavy $\eta$-Mnets $\calM$ of $(X, \calR)$ of size at most  
         \[|\calM| \leq (c_0e)^{d_0}\pth{\frac{4}{1-\lambda}}^{1+\frac{d_0}{\log(1-\Lambda/2)^{-1}}}\cdot\pth{\frac{24d_0}{\eta^2}\psi\pth{\frac{4d_0}{\eta^2},\frac{15d_0}{\eta}}},\]
         where $\Lambda$ can depend on $d$, $\Delta$ and $s$, $c_0$ can depend on $\Delta,s$ and
         $\Lambda,c_0$ are independent of $d_{0}$, $\eta$, $\lambda$ and $n$. $e=2.71\ldots$ is the base of the natural logarithm.
\end{theorem}

\begin{remark}
    The value of $\Lambda$ for a semi-algebraic set systems can be computed from the proof of Theorem~\ref{thm:mnets-semialgebraic-set-systems} 
    in~\cite{DuttaGJM19}, where it is the heaviness constant.
    More specifically for points and halfspaces in $\R^{d}$
    the value of $\Lambda$ is $1/2$, see
    Theorem~\ref{thm:mustafa-ray-14}.  
\end{remark}

        \begin{theorem}[Container bound for semi-algebraic set systems]
         \label{thm:eps-container-semialg-bd}
         If $(X,\calR)$ is a semi-algebraic set system generated by $\Gamma_{d, \Delta, s}$ with VC dimension $d_{0}$ and 
         shallow cell complexity of $(X, \calR^{c})$ is $\psi(\cdot, \cdot)$, then there exists an $\e$-container $\calC$ for $(X, \calR)$ of size at most
         \[
            |\calC| \leq (c_0e)^{d_0}\pth{\frac{4}{\e}}^{1+\frac{d_0}{\log(1-\Lambda/2)^{-1}}}\cdot\pth{\frac{24d_0}{\e^2}\psi\pth{\frac{4d_0}{\e^2},\frac{15d_0}{\e}}},
        \]
         where $c_{0}$ and $\Lambda$ dependent on $d$, $\Delta$, and $s$, and are
         independent of $d_0,\eta,\lambda$ and $n$, and $e=2.71\ldots$ is the base of the natural logarithm.
         \end{theorem}

        \begin{theorem}[Bracket bound for semi-algebraic set systems]
         \label{thm:bracket-bd}
         If $(X,\calR)$ is a semi-algebraic set system generated by $\Gamma_{d, \Delta, s}$ with VC dimension $d_{0}$ and the shallow cell 
         complexities of $(X, \calR)$ and $(X, \calR^{c})$ are $\psi(\cdot, \cdot)$ and $\psi'(\cdot, \cdot)$ respectively, then there 
         exists an $\e$-bracket $\calB$ for $(X, \calR)$ of size at most
         \[
            |\calB| \leq (c_0e)^{d_0}\pth{\frac{8}{\e}}^{1+\frac{d_0}{\log(1-\Lambda/2)^{-1}}} \times \frac{96d_0}{\e^2}\pth{\psi\pth{\frac{8d_0}{\e^2},\frac{30d_0}{\e}} +
                    \psi'\pth{\frac{8d_0}{\e^2},\frac{30d_0}{\e}}},
        \]
         where $c_{0}$ and $\Lambda$ are dependent on $d$, $\Delta$, and $s$, and are
         independent of $d_0,\eta,\lambda$ and $n$, and $e=2.71\ldots$ is the base of the natural logarithm.
         \end{theorem}

\section{Proofs}
\label{sec:proofs}
    We begin with the proof of the Boosting Lemma~\ref{lemma:mnets-1/2-epsilon-Mnet}.

    \begin{proof}[Proof of Lemma~\ref{lemma:mnets-1/2-epsilon-Mnet}]
        Let $\eta' := \min \left\{ \frac{1}{4}, \frac{\eta}{2}\right\}$. 
        Let $\epsilon_{i} := (1+\eta')^{i} \epsilon$ and $\delta_{i} := \eta'\epsilon_{i}$ where $i \in \left\{0, \dots, t\right\}$ and 
        $t = \left\lceil \frac{1}{\log (1+\eta')}\log \frac{1}{\epsilon}\right\rceil$. 
        Let $\calP_{i}$ denote a maximal $(\epsilon_{i}n,\delta_{i}n)$-packing of $\left(X,\calR \right)$. From Theorem~\ref{thm:haussler-packing-lemma}, we 
        have $|\calP_{i}| \leq \left( \frac{cn}{\delta_{i}}\right)^{d}$ where $c$ is an absolute constant. For each $P \in \calP_{i}$, 
        let $\calM(P)$ denote a $\lambda$-heavy $\frac{1}{2}$-Mnet of size at most $\Delta$ for the set system $(P, \calR\mid_{P})$. Let 
        $$
            \calM  := \bigcup_{i=0}^{t} \left( \bigcup_{P \in \calP_{i}} \calM(P) \right).
        $$
        Observe that 
        $$
            \mid\calM\mid = \sum_{i=0}^{t}\left( \sum_{P \in \calP_{i}} \mid\calM(P)\mid \right) \leq \Delta \left(\sum_{i=0}^{t} |\calP_{i}|\right) =
            O\left( \frac{(2c)^{d}\Delta}{\epsilon^{d}}\max\left\{2^{d}, \frac{1}{\eta^{d}}\right\} \right).
        $$
        We will show that $\calM$ is an $\lambda'$-heavy $\epsilon$-Mnet of $(X, \calR)$. Let $R \in \calR$ be such that 
        $\epsilon_{i-1}n \leq |R| < \epsilon_{i}n$. Since $\calP_{i}$ is an $(\epsilon_{i}n,\delta_{i}n)$-packing of $(X, \calR)$, therefore there exists $P \in \calP_{i}$ such that 
        $$
            \mid P \Delta R \mid \; = \; \mid P \setminus R\mid + \mid R\setminus P \mid \; < \; \delta_{i}n.
        $$
        This implies that 
        $$
            |P\cap R| = |R| - |R\setminus P|
            \geq \frac{\epsilon_{i}n}{1+\eta'} - \eta' \epsilon_{i} n
            \geq \epsilon_{i}n \left( 1 - 2 \eta'\right).
        $$
        Note that $1 - 2 \eta' \geq \frac{1}{2}$, and therefore $|P\cap R| \geq |P|/2$.
        Observe that as $\calM(P)$ is a $\lambda$-heavy $\frac{1}{2}$-Mnet of $(P, \calR\mid_{P})$, there exists $Q \in \calM(P)$ such 
        that $Q \subseteq P \cap R$ and $\frac{|Q|}{|R\cap P|} \geq \lambda$. Therefore $Q \subseteq R$, and using the facts 
        $|R| < \epsilon_{i}n$, $|P\cap R|  \geq \epsilon_{i}n \left( 1 - 2 \eta'\right)$ and $\frac{|Q|}{|R\cap P|} \geq \lambda$ we get
        $$
            \frac{|Q|}{|R|} =  \frac{|Q|}{|R\cap P|} \times \frac{|R \cap P|}{|R|} > \lambda (1 - 2 \eta') \geq \lambda (1 - \eta) = \lambda'
        $$
        This completes the proof that $\calM$ is a $\lambda'$-heavy $\epsilon$-Mnet of $(X, \calR)$.
    \end{proof}
    Next we'll prove Lemma~\ref{l:mnet-implies-container} which shows that Mnets and containers have a complementary relationship 
    -- the existence of Mnets in a set system
    implies the existence of containers in the complementary system, and vice versa.

        \begin{proof}[Proof of Lemma~\ref{l:mnet-implies-container}]
            We will denote $\calR^{\leq \delta_{0}n}$ by $\calS$.
            \begin{enumerate}[{\bf (a)}]
                \item 
                    Suppose that $\calM$ is a $\lambda$-heavy $(1-\delta_{0})$-Mnet for $\left( X, \calS^{(c)} \right)$. Observe that for all $R \in \calS$ we have $|R^{c}| \geq (1-\delta_0)n$. Since $\calM$ is a $\lambda$-heavy $(1-\delta_{0})$-Mnet for $\calS^{(c)}$, there exists $M\in \calM$ such that $M \subseteq R^{c}$ and $|M| \geq \lambda (1-\delta_{0})n$. Therefore, we have that $M^{c} \supseteq R$, and using the facts that $|M| \geq \lambda\left( 1-\delta_{0} \right)n$ and $\lambda \in (0,1)$, we have
                    \begin{eqnarray*}
                        |M^{c}\setminus R| = |M^{c}| - |R| = n - |M| - |R|
                        \leq n - \lambda \left( 1 - \delta_{0} \right) n
                        = (1-\lambda + \lambda \delta_{0})n.
                    \end{eqnarray*} 
                    
                \item
                    Suppose $\calC$ is a $(1-\lambda)$-container family for $\left( X, \calS \right)$. For all $R \in \calS$, there exists $C \in \calC$ such that $C \supseteq R$ and $|C \setminus R| \leq (1-\lambda)n$. Therefore, $C^{c} \subseteq R^{c}$, and using the facts that $|C \setminus R| \leq (1-\lambda)n$ and $|R| \leq \delta_{0}n$, we have
                    \begin{equation*}
                        |C^{c}| = n - |C| = n - |C\setminus R| -|R| \geq \left( \lambda -\delta_{0} \right)n.
                    \end{equation*}
            \end{enumerate}
        \end{proof}

Now we come to the proofs of our main theorems, beginning with Theorem~\ref{thm:any-lambda-mnets}. First we show that 
the complementary relation between Mnets and range-containers proved above, can be used recursively to get $\e$-containers 
for small sets in $\calR$.

    \begin{lemma}[Container for small size sets]\label{lem-container-for-small-sets}
        Let $(X,\calR)$  be a set system having VC dimension at most $d_0$, such that there exists a $\Lambda\in (0,1)$, such that 
        given any $\e\in (0,1]$, for any $Y \subset X$, the system $(Y,\calR_{|Y})$ has 
        a $\Lambda$-heavy $\e$-Mnet of size $M^*=M^*_{\Lambda,d_0}(\e)$. Then for any given $0<\rho\leq \e$, there exists a 
        $\rho$-container $\calC_\e$ for $\left( X, \calR^{\leq \e n}\right)$ of size at most 
        \[ M^*\pth{\frac{1}{1+\e/\rho}}^{t_0},\]
   %
        where $t_0:= 1+\frac{\log (1/\e)}{\log(1-\Lambda/2)^{-1}}$.
    \end{lemma}


\begin{proof}
    The idea is, given that ranges are of size at most $\e n$, we construct a $\Lambda$-heavy $(1-\e)$-Mnet $\calM$ for the 
    complement family $\calR^{(c)}:= \left\{ X\setminus R\;|\; R\in \calR \right\}$. We associate with $(X, \calR)$ the set $\calM^{(c)}$, and recurse in the following way:
      Let $\calM = \left\{ C_{1}, \dots, \, C_{l}\right\}$.
    For each $C_{i}$ in the family $\calM$, we recurse on the set system $\left(X_{i}, \calR_{i} \right)$, where 
    \[
        X_{i} = X \setminus C_{i} \;\; \mbox{and} \;\; \calR_{i} = \left\{ R \in \calR \; |\; R \subseteq X_{i} \; \mbox{and} \; |X_{i} \setminus R| > \rho n \right\}.
    \]
    
    Observe that 
    \begin{equation}\label{eqn-reducebyfactor-universesize}
        |X_{i}| = |X\setminus C_{i}| \leq n - |C| \leq n - \Lambda(1-\e) n < \left( 1 - \frac{\Lambda}{2} \right) n, 
    \end{equation}
    where the last two inequalities follow from the facts that $|C| \geq \Lambda (1-\e) n$ and $\e < 1/2$. 
    We construct a $\Lambda$-heavy $(1-\e_{i})$-Mnet $\calM_{i}$ for the set system $\left( X_{i}, \calR_{i}^{(c)} \right)$ where $\e_{i} = \max_{R \in \calR_{i}} \frac{|R|}{|X_{i}|}$, 
    and associate the Mnet $\calM_{i}^{(c)}$ with the set system $(X_{i}, \calR_{i})$.
              
    For any set $\left( Z, \calS \right)$ in this recursion tree with $\calS \neq \emptyset$, observe that for all $R \in \calS$ we have
    \[
        \frac{|R|}{|Z|} = \frac{|R|}{|R| + |Z\setminus R|} = \frac{1}{1+|Z\setminus R|/|R|}< \frac{1}{1+\rho/\e}.
    \]
    The last inequality follows from the fact that $|R| \leq \e n$ and $|Z \setminus R| > \rho n$. Therefore, we have
    \[
        \min_{R \in \calS} \frac{|Z\setminus R|}{|Z|} > 1-\frac{1}{1+\rho/\e} = \frac{1}{1+\e/\rho}.
    \]
    Therefore to reduce the size of $Z\setminus R$ further, we need to find a $\pth{\frac{1}{1+\e/\rho}}$-Mnet
    $\calM_Z$ for $(Z,\calS)^{(c)}$, and recurse by removing a member of $\calM_Z$. By the premise of the Lemma, 
    $\calM_Z$ has size at most \[M^*\pth{\frac{1}{1+\e/\rho}}.\]
    This gives the size of the branching factor at each step of the recursion.
    After every step, the size of the universe is reduced by a factor $\left( 1 - \frac{\Lambda}{2}\right)$, see Equation~\eqref{eqn-reducebyfactor-universesize}.

    Let $i_m$ denote the height of the recursion tree. Therefore, we have 
        $\left( 1 - \frac{\Lambda}{2} \right)^{i_m-1} \geq \e$, 
    that is, the maximum height of the recursion tree satisfies 
    \[
        i_{m} \leq 1+ \frac{\log\pth{\frac{1}{\e}}}{\log\frac{1}{\left( 1 - \frac{\Lambda}{2} \right)}}.
    \]

    To get a container family $\calC$, we do the following: For each node $v$ in the recursion tree, take 
    the union of the Mnet-members $C_i$ which were removed at each node along the path from the root to $v$, 
    and add the complement of this union to the collection $\calC$.
    By our choice of stopping the recursion, the size of the container set is at most $\e n + \rho n \leq 2\e n$.
    The size of the container family is at most the number of root-to-node paths in the recursion tree, i.e. at most the number 
    of nodes, which is bounded by 
    \[ 
        M^*\pth{\frac{1}{1+\e/\rho}}^{1+\frac{\log (1/\e)}{\log(1-\Lambda/2)^{-1}}} = M^*\pth{\frac{1}{1+\e/\rho}}^{t_0}.
    \]
\end{proof}

To extend these small-set containers to containers for the entire set system, we divide the range of possible sizes of 
members of $\calR$ (i.e. $[1,n]$) into a collection of disjoint intervals. We then use an idea similar to that in the proof of the 
Boosting Lemma~\ref{lemma:mnets-1/2-epsilon-Mnet}, to extend the small-set containers, to a container family for 
sets having size in any given interval. This is done in the following lemma.

        \begin{lemma}[Bootstrapping Lemma] 
        \label{l:bstrap-main}
            Given $\e\in (0,1/2]$, $\delta\in (\e,1]$, and a set system $(X,\calR)$ with $|X|=n$ and VC dimension at most $d_0$, 
            and Property $\M$ for some $\Lambda\in (0,1)$, with bound $M^*(\e)=M^*_{\Lambda,d_0}(\e)$
            then there exists a $(1-4\e)$-heavy $\delta$-Mnet for $\left(X,\calR^{[\delta n,(1+\e)\delta n]}\right)$, of size at most 
            $M^*(1/2)^{t_0}\cdot p(\e\delta n,(1+\e)\delta n)$, where $t_0 =t_0(\e,\Lambda) := 1 + \frac{\log 1/\e}{\log (1-\Lambda/2)^{-1}}$.
           
        \end{lemma}

\begin{proof}
Let $\calP$ be a maximal $\e\delta n$-packing of  $\calR^{(\delta n, (1+\e)\delta n]}$. 
    For each $P \in \calP$, let $\calS = \calS(P) := \left\{ A \in \calR^{(\delta n, (1+\e)\delta n]} \;\;\big{|}\;\;|A \Delta P| \leq  \e\delta n\right\}$. 
    Fix $P\in \calP$, and consider the projected set system $(P, \calS|_{P})$. Following our usual notation, let $(\calS_{|P})^{(c)}$ denote the collection of complements 
    of the projected ranges, in $\calS_{|P}$, i.e. 
    \[  \calS_{|P}^{(c)} = \left\{ P\setminus P\cap A \;\;\big{|}\;\; A\in \calR^{(\delta n,(1+\e)\delta n]}, \;\;\&\;\; |A\cap P|\leq \delta\e n \right\}.
    \]
    We claim the following.
    \begin{claim}
    \label{clm:bootstr-pack-in}
       For all $B \in (\calS_{|P})^{(c)}$, $|B| \leq \e'|P|$, where $\e':= \frac{3\e}{2+2\e} \leq 3\e/2$.
    \end{claim}
    
    \begin{proof}
    The proof follows using the fact that $|A|+|P|= |A \Delta P| + 2|A \cap P|$. Since $|A\Delta P|\leq \delta \e n$ and $|A|,|P|\leq (1+\e)\delta n$, 
    we get that $|A \cap P| \geq \delta(1-\e/2)n$. Therefore the ratio $\frac{|P\setminus A|}{|P|}$ can be bounded as  
    \begin{equation}\label{eqn-2}
        \frac{|P\setminus A|}{|P|} =1- \frac{|P\cap A|}{|P|}\leq 1 - \pth{\frac{1-\e/2}{1+\e}}= \frac{3\e}{2+2\e} = \e'. 
    \end{equation}
    \end{proof}
    
    Next, we'll show that an $\e'$-container for $(P,\calS_{|P}^{(c)})$ can yield a $(1-O(\e'))$-heavy Mnet for $\calS$. 
    \begin{claim}
    \label{clm:bootstr-pack-out}
       Let $\calC$ be an $\e'$-container for $(P,\calS_{|P}^{(c)})$. Then $\calC^{(c)}$ is a $(1-4\e)$-heavy $\delta$-Mnet for $\calS$. 
    \end{claim}

    \begin{proof}
       By Claim~\ref{clm:bootstr-pack-in}, each projected range $B\in \calS_{|P}^{(c)}$ satisfies $|B| \leq \e'|P|$. Therefore, 
       by Lemma~\ref{l:mnet-implies-container}~{\bf (b)}, 
       the collection of the complements of $\calC$ in $P$, i.e. $\calC^{(c)}$ is a $(1-2\e')$-heavy $\left( 1 - \e' \right)$-Mnet 
       for the set system $(P, \calS)$.
       Given any $R\in \calS$, i.e. $R\in \calR^{(\delta n,(1+\e)\delta n]}$ and $|R\Delta P|\leq \e\delta n$. By Claim~\ref{clm:bootstr-pack-in}, 
       $|R\cap P| \geq (1-e')\delta n$.
       Then there exists $M \in \calC^{(c)}$ such that $M \subseteq R\cap P$ and $|M|\geq (1-2\e')\delta n$. Therefore, using  
              that $|R| \leq (1+\e)\delta n$ and $|M|\geq (1-2\e')\delta n$, we get 
                \[ 
                    \frac{|M|}{|R|} \geq \frac{|M|}{(1+\e)\delta n}\geq \frac{1-2\e'}{1+\e}  = \frac{1-2\e}{(1+\e)^2} \geq 1-4\e,
                \]
              where in the penultimate step we substituted $\e'=3\e/2(1+\e)$, and in the last inequality we used $\e\leq 1/2$ to 
              bound the Taylor series of $(1-2\e)(1+\e)^{-2}$. 
        Since such an $M\in \calC^{(c)}$ exists for any $R\in \calS$,
        it follows that $\calC^{(c)}$ is a $(1-4\e)$-heavy $\delta$-Mnet for $\calS$. 
    \end{proof}

    Applying Lemma~\ref{lem-container-for-small-sets} (with $\rho=\e$) to the set system $(P, \calS_{|P}^{(c)})$, we get a family $\calC=\calC(P)$ of $\e'$-containers 
    for the set system $(P, \calS_{|P}^{(c)})$ of size 
    \[
    |\calC(P)| = M^*\pth{\frac{1}{2}}^{t_0}. 
    \]
    Therefore, Claim~\ref{clm:bootstr-pack-out} can be applied to get a $(1-4\e)$-heavy $\delta$-Mnet $\calC(P)^{(c)}$ for $\calS$.
    From Proposition~\ref{propn:nn-max-pack}, for every $A\in \calR^{(\delta n, (1+\e)\delta n]}$, there exists a range $P_{A}\in \calP$, 
    such that $|A \Delta P_A| \leq \delta\e n$. Then clearly $\calR^{(\delta n,(1+\e)\delta n]} = \bigcup_{P\in\calP} \calS(P)$.
    Therefore, taking the union of the Mnets $\calC(P)^{(c)}$ gives a $(1-4\e)$-heavy $\delta$-Mnet $\calM$ for $\calR^{(\delta n,(1+\e)\delta n]}$,
    of size 
    \[
         |\calM| = |\calP|\cdot |\calC(P)| = p(\e\delta n, (1+\e)\delta n)\cdot M^*\pth{\frac{1}{2}}^{t_0}. 
    \]
\end{proof}

Now we can combine the container families obtained in the previous lemma, for each of the intervals, to get a container
family for the entire set system.

\begin{proof}[Proof of Theorem~\ref{thm:any-lambda-mnets}]
     For convenience, let us set $\e_0 := \frac{1-\lambda}{4}$, and recall from the statement of the theorem, that $\delta_k = (1+\e_0)^k\eta$.
     The proof follows from a direct application of Lemma~\ref{l:bstrap-main} to each of the families $\pth{\calR^{[\delta_k n,\delta_{k+1} n)}}_{k\geq 0}$.
     Indeed, by applying Lemma~\ref{l:bstrap-main} with parameters $\e=\e_0$ and $\delta=\delta_k$, we get a $(1-4\e_0)$-heavy, i.e. $\lambda$-heavy $\delta_k$-Mnet 
     $\calM_k$ for $\calR^{[\delta_k n,\delta_{k+1}n)}$, of size  $p(\e_0\delta_k n,\delta_{k+1}n)\cdot M^*(1/2)^{t_0}$. Since the union 
     $\calR^{[\delta_k n,\delta_{k+1}n)}$ is $\calR^{[\eta n,n]}$, therefore we get that 
     the union of the families $\calM_k$ for 
     $k\geq 0$ gives the desired $\lambda$-heavy $\eta$-Mnet $\calM$ for $(X,\calR)$. So $\calM$ has size bounded by 
     \[ |\calM| \leq \pth{\sum_{k\geq 0} |\calM_k|} \leq M^*(1/2)^{t_0}\pth{\sum_{k\geq 0} p(\e_0\delta_k n,\delta_{k+1}n)}.\]
     For the second part of the theorem, if $(X,\calR)$ has shallow cell complexity $\psi(.,.)$ then we get from 
     Theorem~\ref{thm:mnets-semialgebraic-set-systems} that $M^*(1/2)$ can be bounded as  
     \[ M^*(1/2) \leq \frac{cd_0}{1/2}\psi\pth{\frac{8d_0}{1/2},48d_0} \leq 2cd_0\psi\pth{16d_0,48d_0},\]
     where $c$ is independent of $d_0,n$ and $\Lambda$. Let $c_0=\max\{2c,16,48\}$, then $M^*(1/2) \leq f_s(c_0d_0)$, 
     where $f_s(r)$ is the \emph{shatter function} at $r$, i.e. the  
     maximum number of projections of $\calR$ on any $r$-subset of $X$.
     By the Sauer-Shelah Lemma~\ref{l:sau-she-71}, this is at most $(c_0ed_0/d_0)^{d_0} \leq (c_0e)^{d_0}$.
     Therefore 
     \[M^*(1/2)^{t_0} = ((c_0e)^{d_0})^{1+\frac{\log 1/\e_0}{\log(1-\Lambda/2)^{-1}}} \leq (c_0e)^{d_0}\cdot\pth{\frac{1}{\e_0}}^{d_0/\log(1-\Lambda/2)^{-1}}.\]
     Also, $\sum_{k\geq 0} p(\delta_k n,l_k n)$ can be bounded as:
     \begin{eqnarray}
         \sum_{k\geq 0} p(\delta_k n,l_k n) &\leq& \sum_{k\geq 0} \frac{24d_0 n}{\delta_k n}\psi\pth{\frac{4d_0 n}{\delta_k n},\frac{12d_0 l_k n}{\delta_k n}} \nonumber\\
                                            &\leq& \sum_{k\geq 0} \frac{24d_0 }{(1+\e_0)^k\eta^2}\psi\pth{\frac{4d_0 }{(1+\e_0)^k\eta^2},\frac{12d_0 (1+\e_0)}{\eta}} \nonumber\\
                                            &\leq&  \frac{1}{\e_0}\frac{24d_0 }{\eta^2}\psi\pth{\frac{4d_0 }{\eta^2},\frac{12(1+1/4)d_0 }{\eta}}.
     \end{eqnarray}
     Therefore, under the assumption that $\psi(.,.)$ is non-decreasing in both its arguments, we get that 
     \[ \sum_{k\geq 0} p(\delta_k n,l_k n) \leq \frac{24d_0/\e_0}{\eta^2}\psi\pth{\frac{4d_0}{\eta^2},\frac{15d_0}{\eta}} .\]
     Putting everything together, we get 
     \[|\calM| \leq (c_0e)^{d_0}\pth{\frac{4}{1-\lambda}}^{1+\frac{d_0}{\log(1-\Lambda/2)^{-1}}}\cdot\pth{\frac{24d_0}{\eta^2}\psi\pth{\frac{4d_0}{\eta^2},\frac{15d_0}{\eta}}}.\]
\end{proof}

\begin{proof}[Proof of Theorem~\ref{thm:eps-container-bd}]
     The proof idea is identical to that of Theorem~\ref{thm:any-lambda-mnets}, except we'll apply it to the system $(X,\calR^{(c)})$. Observe that 
     $(i)$ each range in $\calR^{[(1-\e)n,n]}$ is contained in $X$ with at most $\e n$ extra elements, and further $(ii)$ for the ranges in $\calR^{[0,\e n]}$,
     we can get an $\e$-container family $\calC_1$ of size $\bar{M}^*(1/2)^{t_0}$. So we only need to construct an $\e$-container 
     family for the remaining ranges, i.e. $\calR^{[\e n,(1-\e)n]}$. To do this, observe that $\calR^{[\e n,(1-\e)n]}$ is the complement family of 
     $((\calR^{(c)})^{[\e n,(1-\e)n]})$, so by Lemma~\ref{l:mnet-implies-container} it suffices to construct $(1-\e)$-heavy $\e$-Mnets for $(\calR^{(c)})^{[\e n,(1-\e)n]}$.
     This can be done by applying Theorem~\ref{thm:any-lambda-mnets} to $(\calR^{(c)})^{[\e n,(1-\e)n]}$ with 
     $\lambda=1-\e$, $\eta=\e$. From Lemma~\ref{l:mnet-implies-container}, the complement system of the obtained Mnet is an $\e$-container 
     $\calC_2$ for $\calR^{[\e n, (1-\e)n]}$. The desired $\e$-container for $\calR$ is then given by 
     $\calC = \{X\} \cup \calC_1 \cup \calC_2$. Therefore $\calC$ has size bounded by 
     \[ |\calC| \leq 1 + \bar{M}^*(1/2)^{t_0} + \bar{M}^*(1/2)^{t_0}\cdot\pth{\sum_{k\geq 0}p_{\calR^{(c)}}(\e\delta_k n,\delta_{k+1}n)},\]
     which is at most $2\bar{M}^*(1/2)^{t_0}\cdot \pth{\sum_{k\geq 0}p_{\calR^{(c)}}(\e\delta_k n,\delta_{k+1}n)}$.

     For set systems having shallow cell complexity $\psi(.,.)$, the bound can be worked out nearly exactly as in the proof of Theorem~\ref{thm:any-lambda-mnets}, by 
     replacing $\lambda$ by $1-\e$ and $\eta$ by $\e$. Thus we get 
     \[|\calC| \leq (c_0e)^{d_0}\pth{\frac{4}{\e}}^{1+\frac{d_0}{\log(1-\Lambda/2)^{-1}}}\cdot\pth{\frac{24d_0}{\e^2}\psi\pth{\frac{4d_0}{\e^2},\frac{15d_0}{\e}}}.\]
\end{proof}

\begin{proof}[Proof of Corollary~\ref{cor:unif-brackets}]
     First, we construct a $\e/2$-heavy $\e$-container $\calM$ for $(X,\calR^{(c)})$, using Theorem~\ref{thm:any-lambda-mnets}. 
     Next, we construct an $\e/2$-container $\calC$ for $(X,\calR)$, using 
     Theorem~\ref{thm:eps-container-bd}. We can get an $\e$-uniform bracket $\calB$ by taking the union of $\calM^{(c)}$ and $\calC$.
     This gives the claimed bound on the bracketing number of $(X,\calR)$.
\end{proof}

\begin{proof}[Proof of Theorem~\ref{thm-points-halfspaces-brackets}]
     The idea is to carefully check the dependence of the bound on the bracketing number in terms of the VC dimension $d_0$, which is $d/2$ (ignoring floor and ceiling 
functions). Using Theorem~\ref{thm:mustafa-ray-14} we get that $M^*(1/2)$ is at most $O(4^d\cdot 2^{d/2}) = O(2^{5d/2})$. Therefore, 
$M^*(1/2)^{t_0} = (\e)^{-\frac{(5d/2)\log_2 2}{\log_2 (4/3)}}$. Next, substituting $d_0=d/2$ in the 
Shallow Packing Bound, as used in the proof of Theorems~\ref{thm:any-lambda-mnets} and~\ref{thm:eps-container-bd}, we get a packing bound of 
\[ \sum_{k\geq 0} p(\delta_k n,l_k n) = O\pth{\frac{24d_0}{\e^2}}\psi\pth{\frac{4d_0}{\e^2},\frac{15d_0}{\e}} \leq O\pth{(24ed_0/d_0\e^2)^{d_0}} \leq O\pth{B/\e)^d},\]
where in the penultimate step we used the Sauer-Shelah lemma~\ref{l:sau-she-71}, and in the final step we took $B$ to be a large constant independent of $d$, and substituted $d_0=d/2$.
Substituting these bounds in Corollary~\ref{cor:unif-brackets}, we get that the $\e$-bracketing number of halfspaces in $\R^d$ i.e. $b_{hs}(\e)$, is bounded by 
$B_1^{d}(\e)^{-2d(1+(5/2)/\log (4/3))} \leq (B_2/\e)^{O(d)}$.
\end{proof}



	\bibliography{ref} 
	\bibliographystyle{alpha}

	

\end{document}